\documentclass{article}
\usepackage{arxiv}
\usepackage{twemojis}
\usepackage{hyperref}       
\usepackage{url}            
\usepackage{booktabs}       
\usepackage{amsfonts}       
\usepackage{nicefrac}       
\usepackage[final,nopatch=footnote]{microtype}    
\usepackage{graphicx}

\usepackage{amsmath}
\usepackage{comment}
\usepackage{xcolor}
\usepackage{alltt}
\usepackage{caption} 

\usepackage{float}
\usepackage{authblk}

\usepackage{listings}
\usepackage[T1]{fontenc}

\usepackage{amssymb}
\usepackage{wrapfig}
\usepackage{multirow}
\usepackage{multicol}
\usepackage{biblatex}
\definecolor{keywordcolor}{rgb}{0.7, 0.1, 0.1}   
\definecolor{tacticcolor}{rgb}{0.0, 0.1, 0.6}    
\definecolor{commentcolor}{rgb}{0.4, 0.4, 0.4}   
\definecolor{symbolcolor}{rgb}{0.0, 0.1, 0.6}    
\definecolor{sortcolor}{rgb}{0.1, 0.5, 0.1}      
\definecolor{attributecolor}{rgb}{0.7, 0.1, 0.1} 
\definecolor{mycolor}{rgb}{0.471,0.318,0.663} 

\lstnewenvironment{code}[1][]%
{
   \noindent\newline
   \minipage{\linewidth} 
   \vspace{0.5\baselineskip}
   \lstset{
 	frame = lrtb, 
 	rulecolor=\color{mycolor},
 	escapeinside={/*!}{!*/},
	language=lean, 
	aboveskip = 5mm,
	belowskip = 3mm,
	xleftmargin=2mm,
	xrightmargin=2mm,
	}
	}
{\endminipage\newline}
\lstnewenvironment{codeLong}[1][]%
{
   \lstset{
 	frame = lrtb, 
 	rulecolor=\color{mycolor},
 	escapeinside={/*!}{!*/},
	language=lean, 
	aboveskip = 5mm,
	belowskip = 3mm,
	xleftmargin=2mm,
	xrightmargin=2mm,
	}
	}
{}

\newcommand{\codeLink}[1]{
  \vspace{-0.3cm}\hfill\href{https://github.com/ATOMSLab/LeanDimensionalAnalysis/blob/main/#1}{(source)}
  }
 \newcommand{\textLink}[1]{\href{https://github.com/ATOMSLab/LeanDimensionalAnalysis/blob/main/#1}{source}}
 \newcommand{\textLinkB}[1]{\href{https://github.com/ATOMSLab/LeanDimensionalAnalysis/blob/main/#1}{(source)}}

\title{Formalizing Dimensional Analysis Using the Lean Theorem Prover}

\author[1]{\normalsize Maxwell P. Bobbin}
\author[1]{\normalsize Colin Jones}
\author[1]{\normalsize John Velkey}
\author[1,2]{\normalsize  Tyler R. Josephson}

\affil[1]{\normalsize Department of Chemical, Biochemical, and Environmental Engineering, University of Maryland Baltimore County, \authorcr
1000 Hilltop Circle, Baltimore, MD 21250
}
\affil[2]{\normalsize Department of Computer Science and Electrical Engineering, University of Maryland Baltimore County, \authorcr
1000 Hilltop Circle, Baltimore, MD 21250}

\renewcommand{\shorttitle}{Formalizing Dimensional Analysis}

\hypersetup{
pdftitle={DimensionalAnalysisLean},
pdfsubject={cs.lo},
pdfauthor={Tyler R. Josephson},
pdfkeywords={Proof assistants, Formal verification, Logic, Dimensional Analysis, Lean, Automation},
hidelinks
}
\date{}  

\begin{document}
\maketitle
\begin{abstract}
Dimensional analysis is fundamental to the formulation and validation of physical laws, ensuring that equations are dimensionally homogeneous and scientifically meaningful.
In this work, we use Lean 4 to formalize the mathematics of dimensional analysis.
We define physical dimensions as mappings from base dimensions to exponents, prove that they form an Abelian group under multiplication, and implement derived dimensions and dimensional homogeneity theorems.
Building on this foundation, we introduce a definition of physical variables that combines numeric values with dimensions, extend the framework to incorporate SI base units and fundamental constants, and implement the Buckingham Pi Theorem.
Finally, we demonstrate the approach on an example: the Lennard-Jones potential, where our framework enforces dimensional consistency and enables formal proofs of physical properties such as zero-energy separation and the force law.
This work establishes a reusable, formally verified framework for dimensional analysis in Lean, providing a foundation for future libraries in formalized science and a pathway toward scientific computing environments with built-in guarantees of dimensional correctness.
\end{abstract}

\begin{figure}[h!]
    \centering
    \includegraphics[width=0.9\linewidth]{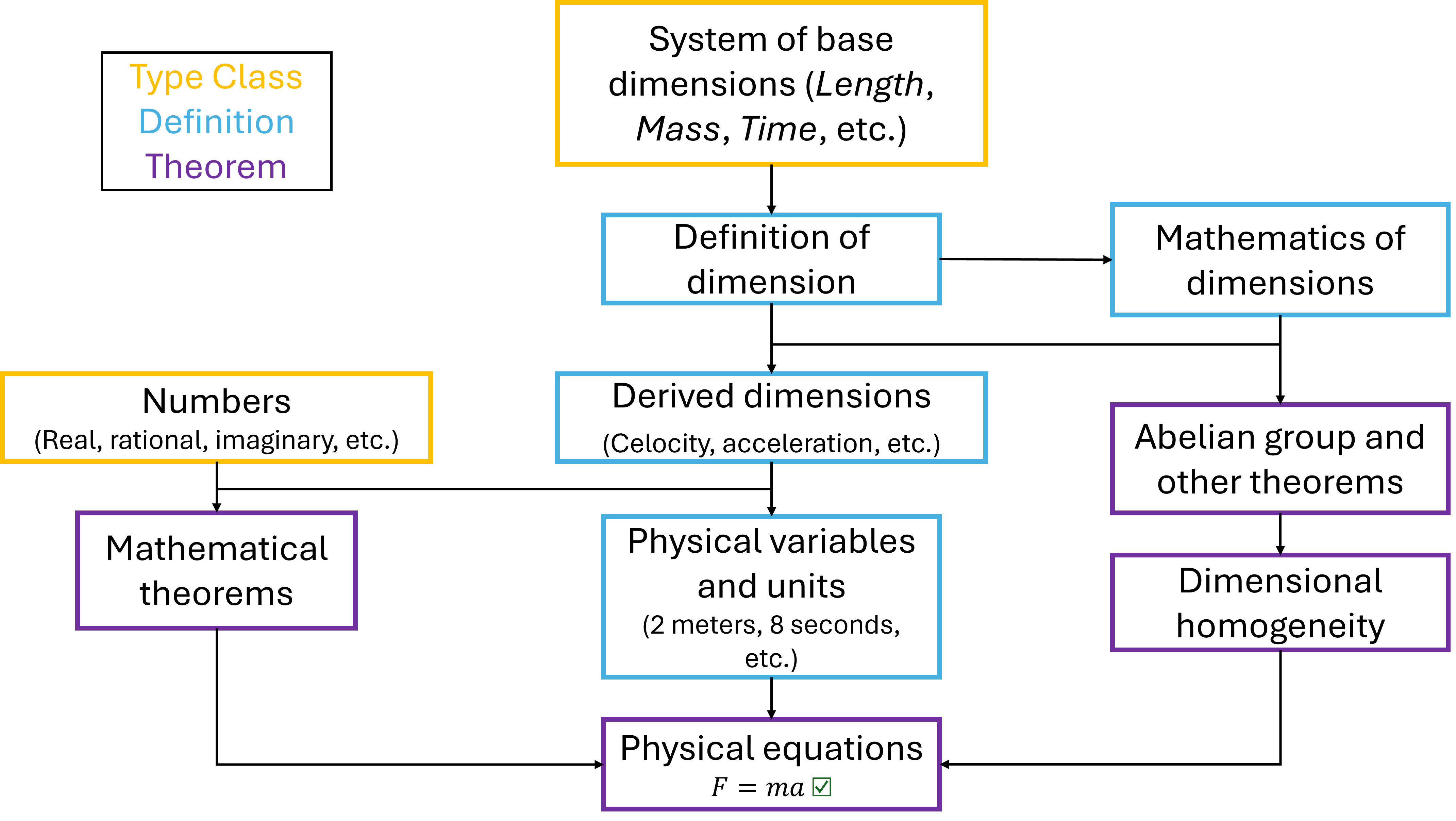}
\end{figure}
\section{Introduction}
\par Dimensions are fundamental to the way we observe, measure, and experiment with the world around us. Physical variables combine numbers and dimensions to describe the real world, and units provide the reference to communicate our observations. All of them are quintessential to engineering and scientific applications. Beyond giving us the ability to verbalize the universe, physical variables contain intrinsic properties that dictate how the formal science of mathematics can explain reality. Joseph Fourier first noted that physical variables can be grouped because they share the same dimension \cite{fourier_theorie_1822}, meaning that they can be compared, and likely describe a common phenomenon. Mathematically speaking, physically valid relations (i.e. relations which describe the real world) can only be those whose dimensions on each side of the relationship are the same. This is the property of dimensional homogeneity and the foundation of scientific mathematics.

Dimensional analysis is the initial tool used to ensure the dimensional homogeneity of formulae \cite{cengel_fluid_2014}, and has been studied in programming languages using type theory since the 1970s \cite{gehani_units_1977,mcbride_type_2021}. Dimensions and units have been coded in common programming languages with type checking, like Fortran\cite{bennich-bjorkman_next_2018}, Ada\cite{gehani_adas_1985}, C++\cite{cmelik_dimensional_1988}, Standard ML\cite{kennedy_programming_1996}, and Haskell \cite{gundry_typechecker_2015}. 
Symbolic programming languages and computer algebra systems (CAS), like SymPy \cite{karam_buckinghampy_2021}, F\#\cite{kennedy_types_2010,owre_automatic_2012}, and more \cite{muradian_diana_1994, albrecht_experimental_2013} have also been used to create programs to alleviate the process, with a focus on the Buckingham Pi theorem. In parallel, the mathematical properties of dimensions and physical variables have been studied \cite{hutter_theoretical_2004,tao_mathematical_2012, autexier_mathematical_2008,sonin_physical_2001}, determining, for instance, that dimensions form an Abelian group for multiplication \cite{fourier_theorie_1822,autexier_mathematical_2008}. However, the code created to implement physical variables and tools, like the Buckingham Pi theorem, has yet to be implemented in a way that formally encompasses the properties of dimensional analysis and the fact that it forms an Abelian group. 

Proof assistants, also known as interactive theorem provers \cite{de_moura_lean_2021,paulson_natural_1986,norrish_thread_2002}, are a type of programming language that would allow the possibility of defining dimensional analysis and verifying the correctness of the definition through formal proofs.
This would result in the same usability as these other programs, but with the added reliability that comes from a formal environment.
In this paper, we present a definition of physical dimensions using the Lean 4 theorem prover and a derivation of the properties of dimensions, such as the formation of an Abelian group under multiplication.
Unlike unit systems in other programs, this implementation is built upon the Lean 4 kernel, ensuring that any theorem written is logically correct, so long as it can be parsed by Lean 4.

Our previous paper has shown the basics of using Lean for scientific applications, where fields such as thermodynamics and kinematics can be formally defined and verified \cite{bobbin_formalizing_2024}. That work introduces formal mathematics and Lean for scientific applications. Tooby-Smith has shown applications of Lean for high energy physics (HEPLean) \cite{tooby-smith_heplean_2025} and digitalizing results of physics (PhysLean) \cite{tooby-smith_formalization_2024}. Ugwuanyi et. al. have shown how Lean can be used for scientific computation with executable Lean code to calculate the potential energy between molecules with the Lennard-Jones potential \cite{ugwuanyi_benchmarking_2025}. We continue that work here by formalizing the mathematics of dimensional analysis in science and engineering, which should find applications both in the development of libraries for formalized science, as well as in the development of scientific computing software with formal guarantees of correctness.

This paper is structured as follows: in Section \ref{sec:overview}, we provide an overview of the nature and mathematics of physical variables and dimensions. Then, in Section \ref{sec:leanimplementation}, we show the implementation of dimensions (\ref{sec:defbasedimension}), the mathematical properties of dimensions (\ref{sec:defmathdimension}), that dimensions form an Abelian group (\ref{sec:abelian}), derived dimensions and dimensional homogeneity (\ref{sec:deriveddimensions}), and physical variables and units (\ref{sec:units}) in Lean. Finally, we illustrate an application in scientific computing, by proving the dimensional homogeneity of the Lennard-Jones function, in Section \ref{sec:applicationtoLJ}.

\section{Methods}
\par 
We used Lean 4, an interactive theorem prover using its mathematical library, Mathlib, version 4.23.0-rc1. Our proofs are hosted on \href{https://github.com/ATOMSLab/LeanDimensionalAnalysis/blob/main/#1}{\underline{GitHub}}. Each code block with a \textit{source} button links to the code line in the GitHub repository.

\section{Overview of Physical Variables and Dimensions}
\label{sec:overview}

A physical quantity is a measure of a system containing a numeric, pure value and a reference unit. A physical variable is a representation of a physical quantity. For example, in $v = 10$ m/s, the velocity, $v$ is the physical variable that represents the idea of an object moving, ``10 m/s'' is the physical quantity, 10 is the value, and m/s is the reference unit. While the physical quantity ``10 m/s'' is tied to its specified units, the physical variable $v$ is more abstract, and has the added bonus of being invariant to the choice of units \cite{boer_history_1995}. The dimensions of $v$ are length/time, these are also invariant to the choice of units.

\subsection{Physical Variables and Their Mathematical Properties}

Formally, we can think of a physical variable, $\mathcal{P}$, as a type containing both a value, $\mathcal{V}$, and a dimension, $\mathcal{D}$, i.e. equation \ref{Physical variable equation}. 
\begin{equation}
    \mathcal{P} = \langle\mathcal{V},\mathcal{D}\rangle
    \label{Physical variable equation}
\end{equation}
When an operator, $\star$, acts on physical variables, it interacts with the value and the dimension of the variable separately, as shown in Eq.~\ref{Physical variable operator}. 
\begin{equation}
    \mathcal{P}_i\star\mathcal{P}_j = \langle\mathcal{V}_i\star\mathcal{V}_j,\mathcal{D}_i\star\mathcal{D}_j\rangle
    \label{Physical variable operator}
\end{equation}
From this, we can understand the mathematical analysis of physical equations to be two-part: an analysis of the values and an analysis of the dimensions. Since Mathlib has developed almost all of the mathematical analysis needed for the scientific analysis of the values, we shall focus on dimensional analysis to define physical variables.  
\subsection{Dimensions and their Mathematical Properties}
The International System of Units, SI units, defines a dimension as a product of International System of Quantities (ISQ) base dimensions raised to a rational\footnote{Rational numbers are a set of numbers that can be represented as the ratio of a natural number and an integer. Examples of rational numbers are $\frac{1}{2}, -2, -\frac{3}{4}$. Real numbers like $\sqrt{2}$ and $\pi$ are excluded as powers in dimensional analysis.} power \cite{newell_international_2019}. There are seven ISQ base dimensions: \textit{Length (L), Time (T), Mass (M), Electric current (I), Temperature ($\theta$), Amount of Substance (N), and Luminous Intensity (J)} \cite{technical_committee_isotc_12_quantities_2022}. Any dimension represented by the ISQ base dimensions can be constructed using Eq.~\ref{ISQ Dimension}. 
\begin{equation}
    \mathcal{D} = L^aT^bM^cI^d\theta ^eN^fJ^g \;\;\;\; {a,b,c,d,e,f,g} \in \mathbb{Q}
    \label{ISQ Dimension}
\end{equation}
Thus, length can be defined by setting $a = 1, b = 0, c = 0, ...$ and velocity can be defined as $a = 1, b = -1, c = 0, d = 0, ...$, and so on for any other dimension. Note that we have just defined two different versions of "length". The first was the base dimension \textit{Length}, and the second was the actual dimension length. A base dimension is used to define a system and construct other dimensions. It will be denoted by capitalizing the first letter and italicizing. Dimensions will be denoted in lower and normal case\footnote{While we will try to avoid it, if a dimension starts a sentence, it will be uppercase, which is why we also italicize base dimensions to avoid confusion.}.
\begin{equation}
    velocity(L) = 1\;\;\;\;
    velocity(T) = -1\;\;\;\;
    velocity(M) = 0
    \label{Velocity Example}
\end{equation}
\begin{wraptable}{r}{7.5cm}
\begin{tabular}{p{1.5cm}p{5cm}}
Operation  & Exponent Manipulation/Result               \\ \hline
$a*b$      & Exponents sum.                             \\
$a/b$      & Exponents subtract.                        \\
$a+a$      & \multirow{2}{*}{Exponents stay the same}   \\
$a-a$      &                                            \\
$Log(a/a)$ & All exponents zero.  
\end{tabular}
\caption{Effects of Operations on Dimension Exponents}
\label{Dimension Math Manipulation}
\end{wraptable}
\par Eq.~\ref{ISQ Dimension} has flaws as a general definition for a dimension. First, it's incomplete because it doesn't account for other base dimensions, like \textit{Currency}, \textit{Number of People}, etc. Second, it is cumbersome and inflexible because it requires us to account for every base dimension, even if our system doesn't use it. Finally, it limits us to rational numbers for exponents, which limits us to rational numbers for powers of physical variables. Therefore, a better definition needs to be able to include new base dimensions easily, allow the user to specify which base dimensions they want to consider, and allow flexibility in the exponent type. 

Therefore, we will define a dimension as a mapping of a base dimension, $\mathcal{B}$, to an exponent, $\mathcal{E}$, Eq.~\ref{Dimension Function}. We also define $\mathcal{E}$ as a type that forms a commutative ring. Since $\mathcal{E}$ forms a commutative ring, the simplest numerical type $\mathcal{E}$ can represent is the integers.
\begin{equation}
    \mathcal{D} = \mathcal{B} \rightarrow \mathcal{E}
    \label{Dimension Function}
\end{equation}
As an example, consider a system with three relevant base dimensions: \textit{Length}, \textit{Time}, and \textit{Mass} (a fundamental system in kinematics). $\mathcal{B}$ has three elements $L$, $T$, and $M$. We can define length as the function that returns $1$ if the base dimension is \textit{Length} and $0$ everywhere else (Eq.~\ref{Length Definition}). Velocity is defined in a similar way (Eq.~\ref{Velocity Definition}).

\begin{equation}
    length(\mathcal{X}) := \begin{cases} 
      1 & \mathcal{X} = L \\
      0 & \mathcal{X} \ne L
   \end{cases}
   \label{Length Definition}
\end{equation}
\begin{equation}
    velocity(\mathcal{X}) := \begin{cases} 
      1 & \mathcal{X} = L \\
      -1 & \mathcal{X} = T \\
      0 & \mathcal{X} \ne L \ne T
   \end{cases}
   \label{Velocity Definition}
\end{equation}
Base dimensions can then be used as inputs into a dimension to determine the respective exponent. Eq.~\ref{Velocity Example} shows an example of indexing the dimension velocity with our system of three base dimensions. 

\begin{wraptable}{r}{8.25cm}
\centering
\begin{tabular}{ll}
\textbf{Associativity} & $\forall\, a\, b\, c, (a*b)*c = a*(b*c)$ \\
\textbf{Identity element} & $\forall\, a, \exists\, e, a*e = e*a = a$ \\
\textbf{Inverse element} & $\forall\, a, \exists\, b, a*b = b*a = e$ \\
\textbf{Commutativity} & $\forall\, a\, b, a*b = b*a$ \\
\end{tabular}

\caption{The mathematical properties of an Abelian group for multiplication. The variables \textit{a} and \textit{b} are any variable from the set for which the Abelian group applies too. The variable \textit{e} is a specific variable called the identity element. The symbol $\forall a$ reads "for all a" and refers to all elements of the set. The symbol $\exists a$ means "there exists an a" and refers to at least one element of the set. }
\label{Abelian Group}
\end{wraptable}
Another, and much more convenient, way to define the velocity dimension would be as the quotient of length and time. Once the math of arithmetic operators is defined in Lean, all derived dimensions will be defined using these operators rather than as step-wise functions. All of these operators involve manipulating the exponent \cite{manner_strong_1986} that is returned when indexed by a base dimension, just like the manipulation commonly shown in Eq.~\ref{ISQ Dimension}. Table \ref{Dimension Math Manipulation} gives an overview of the exponent manipulation for common operators.

As stated above, dimensions form an Abelian group, also known as a commutative group. The properties of an Abelian group are presented in Table \ref{Abelian Group}.
\section{Implementation in Lean}
\label{sec:leanimplementation}
Now we shall turn our attention to defining physical variables and dimensions in Lean. We start with dimensions, which involves defining: how operators act on dimensions, integrating them with Lean's type classes, proving dimensions form an abelian group, creating derived dimensions and dimensional homogeneity theorems, and defining the Buckingham Pi Theorem. Then, we use our definition of dimensions to create physical variables and units, which uses the math defined for dimensions. 

\subsection{Definition of Dimensions and Base Dimensions}
\label{sec:defbasedimension}
The dimension type is defined in Lean based on Equation~\ref{Dimension Function}. It takes two parameters: $B$, representing the system of base dimensions, and $E$, the type used for exponents. The type dimension is then a mapping from base dimensions to exponents:
\begin{code}
/*!\codeLink{DimensionalAnalysis/Basic.lean\#L61}!*/
/-- A dimension is a mapping from each base dimension to the exponent. -/
def dimension (B : Type u) (E : Type v) [CommRing E] := B → E
\end{code}
By leaving $B$ abstract, we allow the same formalism to apply to various systems of base dimensions (e.g., ISQ, mechanical, etc.). For example, consider the kinematic system mentioned previously. This can be defined in Lean as:
\begin{code}
inductive KinematicSystem
| Length | Time | Mass |
\end{code}
Here, inductive is Lean's keyword for defining inductive data types. \textit{KinematicSystem} introduces three base dimensions: \textit{Length}, \textit{Time}, and \textit{Mass}. We could also define a second base dimension system to just consider space and time:
\begin{code}
inductive SpatialTemporalSystem
| Length2 | Time2 |
\end{code}
This system includes elements Length2 and Time2, which correspond to Length and Time in the kinematic system. We intentionally name them differently to emphasize that Lean treats these as distinct types, even though they conceptually represent the same base dimensions. As a result, Lean does not assume any connection between Length and Length2. To reconcile different systems referring to the same physical base dimension, we introduce type classes. To represent the existence of a base dimension \textit{Length}, we define a class HasBaseLength as follows:
\begin{code}
/*!\codeLink{DimensionalAnalysis/Basic.lean\#L16}!*/
/-- Type class for base Length-/
class HasBaseLength (B : Type u) where
[dec : DecidableEq B]
Length : B
\end{code}
This class asserts two things: that equality between elements of the base dimension type $B$ is decidable, and that $B$ includes an element \textit{Length}. By defining instances of this class, we can relate different systems to the same conceptual base dimension:
\begin{code}
/-- Base Length instance for Kinematic System -/
instance : HasBaseLength KinematicSystem :=
{ dec := KinematicSystem.DecidableEq, Length := KinematicSystem.Length }

/-- Base Length instance for Spatial-Temporal System -/
instance : HasBaseLength SpatialTemporalSystem :=
{ dec := SpatialTemporalSystem.DecidableEq, Length := SpatialTemporalSystem.Length2 }
\end{code}
Now, both systems are unified under the shared concept of a \textit{Length} base dimension. This design allows us to write generic code and theorems over arbitrary systems, as long as they satisfy the relevant type class constraints. 
Our current implementation defines a class for all seven ISQ base dimensions, and a currency base dimension to illustrate other potential base dimensions. 
\begin{code}
/*!\codeLink{DimensionalAnalysis/Basic.lean\#L12}!*/
/-- Seven base dimensions from ISQ -/
-- Length defined above
class HasBaseTime (E : Type u) where
  [dec : DecidableEq B]
  Time : E

class HasBaseMass (B : Type u) where
  [dec : DecidableEq B]
  Mass : B

class HasBaseAmount (B : Type u) where
  [dec : DecidableEq B]
  Amount : B

class HasBaseCurrent (B : Type u) where
  [dec : DecidableEq B]
  Current : B

class HasBaseTemperature (B : Type u) where
  [dec : DecidableEq B]
  Temperature : B

class HasBaseLuminosity (B : Type u) where
  [dec : DecidableEq B]
  Luminosity : B

/-- Base dimension for Currency -/
class HasBaseCurrency (B : Type u) where
  [dec : DecidableEq B]
  Currency : B
\end{code}
\subsection{Defining the Mathematics of Dimensions}
\label{sec:defmathdimension}
Multiplication and division are defined as a function that takes in two elements and outputs an element of the same type. Then, the \textit{instance} command is used to globally unify the definition with the respective class. This makes sure that, across Lean, all theorems are talking about the same addition, same multiplication, etc. It also allows us to access the symbols used for these operations ($+,-,*,/,etc$). Multiplication and division for dimensions are defined in Lean as:
\begin{code}
/*!\codeLink{DimensionalAnalysis/Basic.lean\#L93}!*/
/-- Definition of multiplication for dimensions -/
protected def mul {B : Type u} {E : Type v} [CommRing E] : dimension B E → dimension B E → dimension B E
| a, b => fun i => a i + b i
\end{code}
\begin{code}
/*!\codeLink{DimensionalAnalysis/Basic.lean\#L95}!*/
/-- Definition of division for dimensions -/
protected def div {B : Type u} {E : Type v} [CommRing E] : dimension B E → dimension B E → dimension B E
| a, b => fun i => a i - b i
\end{code}
The definition uses \textit{fun} to construct a new dimension by indexing through each base dimension of the input dimensions and adding or subtracting the exponent value for each base dimension. Raising a dimension to a power is defined as a function that takes in a dimension and a value (the value of the power) and outputs a dimension.
\begin{code}
/*!\codeLink{DimensionalAnalysis/Basic.lean\#L100}!*/
/-- Definition of powers for dimensions -/
protected def pow {E E2} [CommRing E] [SMul E2 E]: dimension B E → E2 → dimension B E
| a, n => fun i => n • (a i)
\end{code}
Even though the math has been defined, writing theorems would be cumbersome because we don't have access to the mathematical operators (which will make it easier to read our code and also allow the tactics in Lean to use the definitions). To access the operator symbols, the math that was defined needs to be globally harmonized with its respective classes. This is the same idea used for the base dimension classes. This ensures that, across Lean, all theorems are talking about the same operators. Thus, general theorems about operators can be written and applied to specific cases, like real numbers. For multiplication and division, this looks like:
\begin{code}
/*!\codeLink{DimensionalAnalysis/Basic.lean\#L113}!*/
/-- Unifying multiplication and division definitions with respective type class -/
instance {B : Type u} {E : Type v} [CommRing E] : Mul (dimension B E) := ⟨dimension.mul⟩
instance {B : Type u} {E : Type v} [CommRing E] : Div (dimension B E) := ⟨dimension.div⟩
\end{code}
The rest of the operators are instantiated in the same way. We also implemented differentiation, and describe that in detail in the Supporting Information Section S\ref{Derivative Section}.

\par Defining addition and subtraction takes special care when it comes to dimensional analysis. Two dimensions can be added (or subtracted) only if they are the same dimension. The result is the same dimension. When two scalars that are the same are added together, the result is twice the original number, i.e. \( a + a = 2a\). However, for dimensional analysis, the result is just the original dimension, i.e. \((a : dimension) + a = a\). In the same way, the subtraction of two dimensions should yield the same dimension, \((a : dimension) - a = a\), instead of zero. Addition, like multiplication, is defined as a function that takes in two dimensions and outputs a dimension. However, there is no manipulation of exponents. The definition of addition in Lean is achieved using \textit{Classical.epsilon}, which is the Hilbert epsilon function. Bell \cite{bell_hilberts_1993} and Wirth \cite{wirth_hilberts_2008} both give detailed accounts of the epsilon function, its relation to the axiom of choice, and formal proofs. This gives a formal way of saying \textit{if}  \( \, a = b, \, a + b = a\). In Lean, the definition of addition looks like:
\begin{code}
/*!\codeLink{DimensionalAnalysis/Basic.lean\#L87}!*/
/-- Definition of addition for dimensions-/
protected noncomputable def add {B : Type u} {E : Type v} [CommRing E] : dimension B E → dimension B E → dimension B E := Classical.epsilon \$ fun f => ∀ a b, a = b → f a b = a
\end{code}
The \textit{noncomputable} tag is used to signify that the definition cannot be compiled by Lean for the use of the \textit{\#eval} command, which is a command to evaluate objects. For instance, \textit{\#eval 2 + 2} would return \textit{4} in the Lean infoviewer. This definition is then unified with the addition class so the \textit{+} operator can be used. Substitution is defined in the exact same way and unified with the substitution class. 

\subsection{Proving Dimensions Form an Abelian Group}
\label{sec:abelian}

\par Now, with the math defined for dimensions, we can prove that the Abelian group properties hold. In Lean, the Abelian group, called \textit{CommGroup}, is defined below (note that the numbers are used for the caption to explain each line, but do not actually appear in the Lean code):
\begin{code}
/-- Type class for commutative groups in Mathlib -/
class CommGroup (G : Type u) : Type u
(1)  mul : G → G → G
(2)  mul_assoc (a b c : G) : a * b * c = a * (b * c)
(3)  one : G
(4)  one_mul (a : G) : 1 * a = a
(5)  mul_one (a : G) : a * 1 = a
(6)  npow : ℕ → G → G
(7)  npow_zero (x : G) : Monoid.npow 0 x = 1
(8)  npow_succ (n : ℕ) (x : G) : Monoid.npow (n + 1) x = Monoid.npow n x * x
(9)  inv : G → G
(10) div : G → G → G
(11) div_eq_mul_inv (a b : G) : a / b = a * b⁻¹
(12) zpow : ℤ → G → G
(13) zpow_zero' (a : G) : DivInvMonoid.zpow 0 a = 1
(14) zpow_succ' (n : ℕ) (a : G) : DivInvMonoid.zpow (↑n.succ) a = DivInvMonoid.zpow (↑n) a * a
(15) zpow_neg' (n : ℕ) (a : G) : DivInvMonoid.zpow (Int.negSucc n) a = (DivInvMonoid.zpow (↑n.succ) a)⁻¹
(16) inv_mul_cancel (a : G) : a⁻¹ * a = 1
(17) mul_comm (a b : G) : a * b = b * a
\end{code}

\setlength{\leftskip}{1.5cm}

\setlength{\rightskip}{1.5 cm}
{\small (1) The multiplication operator (defined as a function that takes in two elements and outputs an element). (2) The fact that multiplication is associative (a*b)*c=a*(b*c). (3) The identity element (it is called one, because one is most commonly the identity element for practical representations of numbers). (4) \& (5) the identity element section from the Abelian group definition (Table \ref{Abelian Group}). (6) The natural power operator. (7) The fact that \(x^0=1\). (8) The fact that \(x^{n+1}=x*x^n\). (9) The inverse operator. (10) The division operator. (11) The fact that \(a/b = a*b^{-1}\). (12) The Integer power operator. (13) The same as (7), but for an integer. (14) The same as (8) but for an integer. (15) The fact that \(a^{-n} = (a^n)^{-1}\). (16) The inverse element from the Abelian group definition (Table \ref{Abelian Group}). (17) The commutativity property of the Abelian group definition (Table \ref{Abelian Group}).} 

\setlength{\leftskip}{0 cm}

\setlength{\rightskip}{0 cm}

\par At first glance, \textit{CommGroup} appears to be more in-depth than the Abelian group. However, it does not define anything outside of the Abelian group. Instead, it has to talk about each case (the four parts to the table, plus division to make programming easier \footnote{The division and the inverse operator go hand in hand, so it is natural to bring in division when the inverse operation is talked about. That is why, even though Table \ref{Abelian Group} does not explicitly mention the division operator, \textit{CommGroup} still talks about it.}and the operators needed (multiplication, division, inverse, npow, and zpow, along with the identity element). Next, we will walk through the proofs done in Lean to show each of these properties. Finally, we show the instance command that proves to Lean that \textit{dimension} forms an Abelian group.

\par The first property we show is that multiplication is commutative (17), (\(a*b=b*a\)). In Lean, this theorem looks like:
\begin{code}
/*!\codeLink{DimensionalAnalysis/Basic.lean\#L188}!*/
/-- Theorem proving that multiplication of dimensions is commutative -/
protected theorem mul_comm {B : Type u} {E : Type v} [CommRing E] (a b : dimension B E) : a * b = b * a := by
  simp only [mul_def']
  funext
  rw [add_comm]
\end{code}
Next, multiplication by the identity element (4) is shown. The other version  (5) can be shown by using the \textit{mul\_comm} theorem that was just proven. This theorem answers the question of what the identity element for dimensions is. Since multiplication involves adding the values of the exponents of the two dimensions being multiplied, multiplying by a dimensionless dimension results in adding zero to all the values, which preserves the original dimension. In Lean, this looks like this:
\begin{code}
/*!\codeLink{DimensionalAnalysis/Basic.lean\#L206}!*/
/-- Theorem proving that one (dimensionless) multiplied by a dimension equals 
the original dimension -/
protected theorem one_mul {B : Type u} {E : Type v} [CommRing E] (a : dimension B E) : 1 * a = a := by simp only [one_eq_dimensionless,
  dimensionless_def', Function.const_zero, mul_def', Pi.zero_apply, zero_add]
\end{code}
Next, the associativity of multiplication (2) is shown: 
\begin{code}
/*!\codeLink{DimensionalAnalysis/Basic.lean\#L198}!*/
/-- Theorem proving that multiplication of dimensions is associative -/
protected theorem mul_assoc {B : Type u} {E : Type v} [CommRing E] (a b c : dimension B E) : a * b * c = a * (b * c) := by
  simp only [mul_def']
  funext
  rw [add_assoc]
\end{code}
\par The final property we show is the inverse element (16) for dimensions. Another part that is added in the \textit{CommGroup} definition is the relationship between division and multiplying by the inverse (\(a/b = a*b^{-1}\)). Both of those are shown below. Note that in the \textit{mul\_left\_inv} proof, the number 1 is used in place of \textit{dimensionless}. The number 1 is the identity element operator just like * is the multiplication operator, and comes from unifying \textit{dimensionless} with \textit{one}.
\begin{code}
/*!\codeLink{DimensionalAnalysis/Basic.lean\#L214}!*/
/-- Theorem proving that the inverse of a dimension multiplied by the 
same dimension yields dimensionless (one) -/
protected theorem mul_left_inv {B : Type u} {E : Type v} [CommRing E] (a : dimension B E) : a⁻¹ * a = 1 := by
  simp
  funext
  simp
\end{code}
\begin{code}
/*!\codeLink{DimensionalAnalysis/Basic.lean\#L209}!*/
/-- Theorem proving the relation between division and multiplication by an inverse -/
protected theorem div_eq_mul_inv {B : Type u} {E : Type v} [CommRing E] (a b : dimension B E) : a / b = a * b⁻¹ := by
  simp
  funext
  rw [sub_eq_add_neg]
\end{code}
In all of these proofs, tactics like \textit{simp} and \textit{funext} were used extensively along with a couple of other lemmas, whose proofs were not shown, but available on GitHub. The ability to use tactics to simplify the proof process is a result of deriving a large set of helper lemmas attached to the \textit{simp} tactic that automate the tedious process of reverting back to the base definition of dimension and applying proofs to function mappings.
\par The final step is to use the \textit{instance} command to prove to Lean that dimensions form an Abelian group. To do this, we must give the proof for each part of the Abelian group. Since we have already proven the individual theorems, we just have to reference the theorem. In Lean: 
\begin{code}
/*!\codeLink{DimensionalAnalysis/Basic.lean\#L234}!*/
/-- Instance proving that dimensions form a commutative (abelian) group -/
instance {B : Type u} {E : Type v} [CommRing E] : CommGroup (dimension B E) where
  mul := dimension.mul
  div := dimension.div
  inv a := dimension.pow a (-1)
  mul_assoc := dimension.mul_assoc
  one := dimensionless B E
  npow n a := dimension.pow a ↑n
  zpow z a:= dimension.pow a ↑z
  one_mul := dimension.one_mul
  mul_one := dimension.mul_one
  mul_comm := dimension.mul_comm
  div_eq_mul_inv a := dimension.div_eq_mul_inv a
  inv_mul_cancel a := dimension.mul_left_inv a
  npow_zero := by intro x; funext x; simp
  npow_succ n a := by simp; funext x; rw [add_one_mul]
  zpow_neg' _ _ := by simp; rename_i x1 x2; funext x3; rw [← neg_add,neg_mul,add_comm]
  zpow_succ' _ _ := by simp; rename_i x1 x2; funext; rw [add_one_mul]
  zpow_zero' := by intro x; funext x; simp
\end{code}

\subsection{Derived Dimensions and Dimensional Homogeneity Theorems}
\label{sec:deriveddimensions}
Now that all the math has been defined and unified with Lean's class system, regular dimensions can be defined, and theorems about the dimensional homogeneity of equations can be written and easily proved. As was mentioned above, the dimension \textit{length}, Eq. \ref{Length Definition}, is defined as a function that evaluates to $1$ at \textit{Length} and $0$ everywhere else. In Lean, this looks like:
\begin{code}
/*!\codeLink{DimensionalAnalysis/Dimensions.lean\#L12}!*/
/-- The dimension length -/
def length (B : Type u) (E : Type v) [CommRing E] [HasBaseLength B] : dimension B E := Pi.single HasBaseLength.Length 1
\end{code}
The \textit{Pi.single} creates a function which is $1$ at the base dimension \textit{Length} and $0$ everywhere else. \textit{Pi} refers to a Pi type, which is a dependent function type. In this case, we have a standard function type, which is a Pi type constrained so the function has the same type output regardless of the input parameter (i.e., the function outputs a type $E$ regardless of the element of $B$ passed). The \textit{[HasBaseLength $B$]} part requires that the system "has \textit{Length}" as one of its base dimensions. The system could have other base dimensions, but the only important one for this definition is \textit{Length}. The command \textit{Pi.single HasBaseLength.Length 1} creates a function that is $1$ at \textit{HasBaseLength.Length} (the base dimension \textit{Length}) and zero everywhere else. This approach is both flexible and expandable, allowing for new systems or base dimensions to be added in a modular fashion without modifying existing logic or definitions.
\par The dimension \textit{time} can be defined in the same way, except it requires that $B$ "has \textit{Time}" instead of \textit{Length}.
\begin{code}
/*!\codeLink{DimensionalAnalysis/Dimensions.lean\#L13}!*/
/-- The dimension time -/
def time (B : Type u) (E : Type v) [CommRing E] [HasBaseTime B] : dimension B E := Pi.single HasBaseTime.Time 1
\end{code}
\par More advanced dimensions can be defined in a much more familiar way. While we could use the \textit{Pi} function to continue defining dimensions, we can instead use the primary dimensions just defined and the math defined in the previous section. The dimensions \textit{velocity} and \textit{acceleration} can be defined as:
\begin{code}
/*!\codeLink{DimensionalAnalysis/Dimensions.lean\#L30}!*/
/-- The dimension velocity -/
abbrev velocity (B : Type u) (E : Type v) [CommRing E] [HasBaseLength B] [HasBaseTime B] := length B E / time B E
/-- The dimension acceleration -/
abbrev acceleration (B : Type u) (E : Type v) [CommRing E] [HasBaseLength B] [HasBaseTime B] := length B E / ((time B E) ^ 2)
\end{code}
In this example, $B$ needs to have both \textit{Time} and \textit{Length} since it references the length and time definition. Since the division of two dimensions is defined as subtracting their exponents, this creates a function which is $1$ at \textit{Length}, $-1$ at \textit{Time}, and zero everywhere else. \textit{abbrev} is syntactic sugar for marking a definition as reducible. We choose to make constructed dimensions reducible so Lean's type checker can automatically look inside the definition. This makes it easier to prove dimension homogeneity theorems. Another example is the Reynolds number:
\begin{code}
/*!\codeLink{DimensionalAnalysis/Dimensions.lean\#L47}!*/
/-- Dimension of the Reynolds number -/
abbrev reynolds_number (B : Type u) (E : Type v) [CommRing E] [HasBaseLength B] [HasBaseTime B] [HasBaseMass B] := mass_density B E * velocity B E * length B E / dynamic_viscocity B E
\end{code}
\par Finally, theorems about the dimensional homogeneity of equations can be written. For instance, this is what a theorem showing that acceleration is dimensionally homogeneous to velocity divided by time looks like in Lean. 
\begin{code}
/*!\codeLink{DimensionalAnalysis/DimensionalHomogeneity.lean\#L7}!*/
/-- Theorem proving the relation between the dimension of acceleration, velocity, 
and time. -/
theorem accel_eq_vel_div_time {B E} [CommRing E][HasBaseLength B] [HasBaseTime B] : acceleration B E = velocity B E / time B E := by rw[acceleration,velocity,pow_two,div_div]
\end{code}
Another example is shown below, showing that the Reynolds number is dimensionless:
\begin{code}
/*!\codeLink{DimensionalAnalysis/DimensionalHomogeneity.lean\#L10}!*/
/-- Theorem proving that the Reynolds number is dimensionless -/
theorem reynolds_eq_dimless (B : Type u) (E : Type v) [CommRing E] [HasBaseLength B] [HasBaseTime B] [HasBaseMass B] : reynolds_number B E = dimensionless B E := by
  rw [reynolds_number,mass_density,volume,velocity,dynamic_viscocity, ← one_eq_dimensionless, div_eq_one]
  rw [mul_assoc,mul_comm (length B/time B),mul_div,pow_three,
  ← mul_div_assoc,mul_comm,← mul_div_assoc,mul_comm _ (length B E * length B E), mul_div_mul_comm, 
  ← div_one (length B E * length B E),div_div_div_cancel_left,div_one,one_mul,div_div]
\end{code}

\subsection{Physical Variables and Units}
\label{sec:units}
With dimensions fully defined, we can build on this to define physical variables. This is done as a graded structure over a dimension with a field for the value of the measurement. 
\begin{code}
/*!\codeLink{PhysicalVariables/Basic.Lean\#L5}!*/
/-- Definition of a physical variable -/
structure PhysicalVariable {B : Type u} {V : Type v} [Field V] (dim : dimension B V) where
(value : V)
\end{code}
Here, $B$ is the type representing the system of base dimensions, $V$ is the type of the measured value, and $d$ is the dimension the measurement is made on. We also require the value type and exponent type to be the same, as it is easier to read and doesn't reduce the applicability. With the graded structure, we can encode the dimension manipulation of an operator directly into the type. For multiplication this looks like:
\begin{code}
/*!\codeLink{PhysicalVariables/Basic.Lean\#L16}!*/
/-- Definition of multiplication for physical variables -/
protected def Mul {B : Type u} {V : Type v} [Field V] {d1 d2 : dimension B V}:
PhysicalVariable d1 →  PhysicalVariable d2 → PhysicalVariable (d1*d2)
| a,b => PhysicalVariable.mk (a.value*b.value)
\end{code}
This definition makes use of the multiplication defined for the value type and dimension. Therefore, an operator can be defined for a physical variable as long as the operator exists for both the value type and dimension. For addition, we can directly encode the requirement of dimensional homogeneity by writing:
\begin{code}
/*!\codeLink{PhysicalVariables/Basic.Lean\#L35}!*/
/-- Definition of addition for physical variables -/
protected def Add {B : Type u} {V : Type v} [Field V] {d : dimension B V} :
  PhysicalVariable d → PhysicalVariable d → PhysicalVariable d
| a, b => ⟨a.value + b.value⟩
\end{code}
Unlike with dimensions, where we had to use the epsilon operator to require the dimensions to be the same to add, with a graded structure for physical variables, we can require that addition only holds for physical variables with the same dimension. These are both harmonized with the type class for its respective operator and this is also done for division and subtraction. However, we cannot unify our power definition with the power type class. Our power definition is:
\begin{code}
/*!\codeLink{PhysicalVariables/Basic.Lean\#L50}!*/
/-- Definition of powers for physical variables -/
protected def Pow {B : Type u} {V1 : Type v} {V2} [Field V1] [HPow V1 V2 V1] [SMul V2 V1] {d : dimension B V1} (a : PhysicalVariable d) (n : V2):
(PhysicalVariable ((d^n) : dimension B V1) ) := ⟨a.value^n⟩
\end{code}

This cannot be written in a full function form, because we have to know the power $n$ to know the output dimension. To write $a^b$, we would write \textit{a.Pow b}. This doesn't reduce the usability of the code, just the presentation of the code.

With this definition of physical variables, we run into a slight problem when writing physical variable equations on mixed dimensions. For example, if Newton's second law $F=ma$ was written using this formulation, we would get an error because the dimension force is not definitionally equal to the dimension mass times acceleration. However, it is prepositionally equal. This means Lean cannot automatically do a type class inference on this equation and throws an error. To get around this, we define a cast function, inspired by ecrybe (see acknowledgments), which allows for converting propositionally equal dimensions. 

\begin{code}
/*!\codeLink{PhysicalVariables/Basic.Lean\#L10}!*/
/-- Cast function, with up-arrow notation, to convert prepositionally equal dimensions -/
protected def cast {B : Type u} {V : Type v} [Field V] {d1 d2 : dimension B V} (Q : PhysicalVariable d1) (_ : d1=d2 := by evalAutoDim) :
PhysicalVariable d2 := ⟨Q.value⟩

prefix:100 (priority := high) "↑" => PhysicalVariable.cast
\end{code}

This makes use of an empty/placeholder premise that $d1=d2$ and the \textit{:= by evalAutoDim} tells Lean to run our custom tactic \textit{evalAutoDim} to try and prove the dimensional homogeneity. The tactic is built as:
\begin{code}
/*!\codeLink{DimensionalAnalysis/DimensionalHomogeneity.lean\#L16}!*/
/-- Tactic to automatically prove dimensional homogeneity. -/
macro "evalAutoDim" : tactic =>
  `(tactic|
    (first | rfl
           | try rw [mul_one,one_mul,mul_comm,one_eq_dimensionless]
             try simp
             try funext
             try module
             try ring_nf
             try field_simp
             try simp
             try rfl

    ))
\end{code}
The user could also supply a proof directly of the dimensional homogeneity if the tactic cannot close the goal, but, for all the cases we tested, we found the tactic to be strong enough to close the goal and it can be easily expanded as new cases arise. 

Even though we can write physical equations that ensure dimensional homogeneity, there is one more thing missing for scientific computation: units, which provide a scale for a measurement. Defining units requires defining what "1" is, since this is the reference used to build measurements. We will make use of the SI units definition, as this is the basis of scientific calibration. The definition of the SI base units from the International Bureau of Weights and Measures is outlined in Table \ref{SI Base Units}.

\begin{table}[h!]
\caption{Definition of the seven base SI units as of 2019 \cite{noauthor_international_2019}.}
\begin{tabular}{p{2.5cm}|p{13cm}}
SI Unit       & Definition \\ \hline
Second (s)    & Defined as $9,192,631,770$ oscillations of the unperturbed ground-state hyperfine transition frequency of the Caesium (Cesium)-133 atom. \\
Meter (m)     & Defined as setting the speed of light in vacuum to be $299,792,458$ in units of m/s, using the definition of second. \\
Kilogram (kg) & Defined as setting Planck's constant to be $6.62607015\times10^{-34}$ in units of kg m$^2$/s, using the definition of meter and second. \\
Ampere (A)    & Defined as setting the elementary charge to be $1.602176634\times10^{-19}$ in units of A s, using the definition of the second. \\
Kelvin (K)    & Defined as setting the Boltzmann constant to be $1.380649\times10^{-23}$ in units of kg m$^2$/(s$^2$K), using the definition of kilogram, meter, and second. \\
Mole (mol)    & Defined as $6.02214076\times 10^{23} $elementary entities. \\
Candella (cd) & Defined as setting the luminous efficacy of monochromatic radiation at a frequency of $540$ THz to be $683$ in units of cd sr/(kg m$^2$s$^3$), using the definition of kilogram, meter, and second.
\end{tabular}
\label{SI Base Units}
\end{table}
The base dimensions are implemented in Lean below. Starting with time, the duration of the ground state hyperfine oscillation of Caesium-133 is defined as one time. From there, a second is defined as 9,192,631,770 of those oscillations. 
\begin{code}
/*!\codeLink{PhysicalVariables/Basic.Lean\#L171}!*/
/-- Definition of the unit of time for a single caesium-133 oscillation -/
def casesium133GroundStateHyperfineOscillationDuration {B : Type u} {V : Type v} [Field V] [HasBaseTime B] :
PhysicalVariable (dimension.time B V) := ⟨1⟩
/-- Definition of the second based on the caesium-133 atom -/
def second (B : Type u) (V : Type v) [Field V] [HasBaseTime B] : PhysicalVariable (dimension.time B V) := 9192631770•casesium133GroundStateHyperfineOscillationDuration

\end{code}
The meter is defined as one length. With the definition of the meter and the second, we can then define the speed of light as exactly 299,792,458 m/s. 
\begin{code}
/*!\codeLink{PhysicalVariables/Basic.Lean\#L176}!*/
/-- Definition of the meter -/
def meter (B : Type u) (V : Type v) [Field V] [HasBaseLength B] : PhysicalVariable (dimension.length B V) := ⟨1⟩

/-- Definition of the speed of light using the meter and second -/
def SpeedOfLight (B : Type u) (V : Type v) [Field V]   [HasBaseLength B] [HasBaseTime B] : PhysicalVariable (dimension.length B V / dimension.time B V) :=
  299792458 • meter B V/second B V
\end{code}
Following the same theme, the remaining five base units are defined as one of their respective dimension and the values of the five corresponding physical constants are fixed with those units.
\begin{code}
/*!\codeLink{PhysicalVariables/Basic.Lean\#L168}!*/
/-- Definition of the kilogram -/
def kilogram (B : Type u) (V : Type v) [Field V] [HasBaseMass B] : PhysicalVariable (dimension.mass B V) := ⟨1⟩
/-- Definition of the ampere -/
def ampere (B : Type u) (V : Type v) [Field V] [HasBaseCurrent B] : PhysicalVariable (dimension.current B V) := ⟨1⟩
/-- Definition of the kelvin -/
def kelvin (B : Type u) (V : Type v) [Field V] [HasBaseTemperature B] : PhysicalVariable (dimension.temperature B V) := ⟨1⟩
/-- Definition of the mole -/
def mole (B : Type u) (V : Type v) [Field V] [HasBaseAmount B] : PhysicalVariable (dimension.amount B V) := ⟨1⟩
/-- Definition of the candela -/
def candela (B : Type u) (V : Type v) [Field V] [HasBaseLuminosity B] : PhysicalVariable (dimension.luminosity B V) := ⟨1⟩
\end{code}
\begin{code}
/*!\codeLink{PhysicalVariables/Basic.Lean\#L191}!*/
/-- Definition of Planck's constant using the meter, kilogram, and second -/
def PlancksConstant (B : Type u) (V : Type v) [Field V] [HasBaseLength B] [HasBaseTime B] [HasBaseMass B] [SMul Float V]:
  PhysicalVariable (dimension.mass B V * dimension.length B V ^ 2 / dimension.time B V) :=
  6.62607015e-34•(kilogram B V * (meter B V).Pow 2 / second B V)
/-- Definition of the Elementary Charge using the ampere and second -/
def ElementaryCharge (B : Type u) (V : Type v) [Field V] [HasBaseCurrent B] [HasBaseTime B] [SMul Float V]:
  PhysicalVariable (dimension.current B V * dimension.time B V) :=
  1.602176634e-19 • (ampere B V * second B V)
/-- Definition of Boltzman's constant using the meter, second, and kelvin -/
def BoltzmannConstant (B : Type u) (V : Type v) [Field V]
  [HasBaseLength B] [HasBaseTime B] [HasBaseTemperature B] [SMul Float V]:
  PhysicalVariable (dimension.length B V ^ 2 / (dimension.time B V ^ 2 * dimension.temperature B V))  :=
  1.380649e-23 • ((meter B V).Pow (2 : ℕ) / ((second B V).Pow 2 * kelvin B V))
/-- Definition of Avogadros Number from the mole -/
def AvogadrosNumber (B : Type u) (V : Type v) [Field V] [HasBaseAmount B] [Pow V V] [SMul Float V]:
  PhysicalVariable ((dimension.amount B V) ^(-1:ℤ)) :=
  6.02214076e23 • (mole B V).Pow (-1:ℤ)
/-- Defnition of the luminous efficacy of 540 THz monochromatic light -/
def MonochromaticRadiation540THz (B : Type u) (V : Type v) [Field V] [Pow V V]
  [HasBaseLength B] [HasBaseTime B] [HasBaseMass B] [HasBaseLuminosity B] :
  PhysicalVariable (dimension.luminosity B V / (dimension.mass B V * dimension.length B V ^ 2 * dimension.time B V ^ 3)) :=
  683 • ↑(candela B V * steradian B V)/ (kilogram B V * (meter  B V).Pow 2 * (second B V).Pow 3)
\end{code}

\section{Discussion: Application to Scientific Computing with the  Lennard Jones Potential}
\label{sec:applicationtoLJ}
Ugwuanyi, et. al. showed how Lean could be used to create a formalized and executable framework to compute molecular interaction energies with the Lennard-Jones potential \cite{ugwuanyi_benchmarking_2025}. Their approach validated the math and the algorithms involved int he calculation, but did not incorporate units or have any checks to ensure dimensional consistency of their equations.
In this section, we show how this code can incorporate our physical variable definition to allow us to also ensure dimensional and unit homogeneity. To do so, we will show how the Lennard-Jones potential can be defined in Lean with physical variables and two theorems about the Lennard-Jones potential. 

The Lennard-Jones potential, Eq. \ref{Lennard-Jones potential}, describes the energy between two particles, where $V$ is the potential energy, $\varepsilon$ is the depth of the potential well (the minimum energy in the interaction), $\sigma$ is the distance where the energy between the two molecules is zero (attractive and repulsive forces are balanced), and $r$ is the distance between the molecules.
\begin{equation}
    \label{Lennard-Jones potential}
    V = 4\varepsilon\left(\left( \frac{\sigma}{r}\right)^{12}-\left( \frac{\sigma}{r}\right)^6\right)
\end{equation}
In Lean, we define this as:
\begin{code}
/*!\codeLink{PhysicalVariables/LennardJones.lean\#L15}!*/
/-- Definition of the Lennard-Jones potential -/
noncomputable def LennardJonesPotentialEnergy {B V} [Field V] [HasBaseLength B] [HasBaseTime B] [HasBaseMass B] (σ : PhysicalVariable (dimension.length B V))
(ε : PhysicalVariable (dimension.energy B V)) (r: PhysicalVariable (dimension.length B V)):
  PhysicalVariable (dimension.energy B V) := 4 • ↑(ε * (↑(σ/r).Pow (12) - (σ/r).Pow 6))
\end{code}
This definition requires a system of base dimensions with \textit{Length}, \textit{Time}, and \textit{Mass}, as well as three parameters. The first two are Lennard-Jones parameters specific to a system and the last is the distance between the molecules. The up arrows are our cast definition. The inner most up arrow converts the dimension of $(\sigma/r)^{12}$ to an equivalent form so it can be subtracted. The second arrow converts the dimension energy times dimensionless to energy. Both of these are automatically proven by our custom tactic.

The first theorem we show is that the Lennard-Jones potential gives zero energy when the radius equals $\sigma$. 
\begin{code}
/*!\codeLink{PhysicalVariables/LennardJones.lean\#L19}!*/
/-- Theorem proving that when the molecules are separated by distance σ, the energy 
is 0 -/
theorem LJ_zero_energy {B V} [Field V] [HasBaseLength B] [HasBaseTime B] [HasBaseMass B] (σ : PhysicalVariable (dimension.length B V))
(ε : PhysicalVariable (dimension.energy B V)) (hσ : σ.value ≠ 0) :
  LennardJonesPotentialEnergy σ ε σ  = ⟨0⟩ := by
/-- rest of proof on GitHub-/
\end{code}
The other theorem we will show is the derivative of the Lennard-Jones potential with respect to the distance between the molecules, which describes the force between the two molecules. This makes use of our definition of the single variable derivative for physical variables (Supporting Information S\ref{Derivative Section}).
\begin{code}
/*!\codeLink{PhysicalVariables/LennardJones.lean\#L43}!*/
/-- Theorem showing the force between two molecules that follow a 
Lennard-Jones potential -/
theorem LJ_deriv {B V} [NontriviallyNormedField V] [HasBaseLength B] [HasBaseTime B] [HasBaseMass B] (σ : PhysicalVariable (dimension.length B V))
(ε : PhysicalVariable (dimension.energy B V)) {r : PhysicalVariable (dimension.length B V)} (hr0 : r.value ≠ 0) :
PhysicalVariable.deriv (LennardJonesPotentialEnergy σ ε) r =  4 • ↑(ε * (-12•↑(σ.Pow 12/r.Pow 13) + 6•σ.Pow 6/r.Pow 7)) := by
/-- rest of proof on GitHub-/
\end{code}

\section{Conclusion}
Here, we have shown how dimensional analysis and physical variables can be defined using Lean. We started by defining dimensions in Lean and showing that they form an Abelian group. We can use this foundation to write theorems about dimensional homogeneity of equations and the implementation of the Buckingham Pi Theorem. Our definition of dimensions can be used to create physical variables and units based on definitions from the International Bureau of Weights and Measures.
Finally, we showed an application of this code to the Lennard-Jones function, highlighting the ability to ensure dimensional consistency within proofs. 
This code should provide the framework necessary to construct physically meaningful equations and perform scientific computations in Lean, with dimensional consistency ensured via type checking.

\section*{Acknowledgements}

We are grateful to the Lean prover community and contributors of Mathlib on whose work this project is built. We especially acknowledge the work of Joseph Tooby-Smith and Alfredo Moriera-Rosa, who created independent formulations of physical variables in varying states of implementation. These can be found \textcolor{blue}{\href{https://github.com/HEPLean/PhysLean/tree/master/PhysLean/Units}{here}} and \textcolor{blue}{\href{https://github.com/ecyrbe/lean-units}{here}}, respectively. Alfredo Moriera-Rosa and Terence Tao provided inspiration and advice in formulating the cast function and tactic to make the graded structure work for physical variables. This material is based on work supported by the National Science Foundation (NSF) CAREER Award \#2236769.

\printbibliography
\newpage

\title{FORMALIZING DIMENSIONAL ANALYSIS USING THE LEAN THEOREM PROVER\\Supplementary Information}
\maketitle
\setcounter{section}{0}
\section{The Buckingham Pi Theorem}
\label{Buckingham Pi Section}
The Buckingham Pi Theorem states that, given a set of dimensions that form a dimensional matrix, we can calculate the number of dimensionless numbers that can be formed and construct those numbers. The dimensionless matrix of a list of \(n\) dimensions with \(k\) base dimensions is a \(k \times n\) matrix where each entry, \((i,j)\), corresponds to the value of the exponent for base dimension \(i\) in variable \(j\). For the set of dimensions: (length, time, and velocity), the dimensional matrix would look like:
\begin{equation}
    \label{Dim Matrix Example 1}
    \begin{bmatrix}
    1 & 0 & 1 \\
    0 & 1 & -1
    \end{bmatrix}
\end{equation}
Since the dimensional matrix must be of a form where the rows correspond to base dimensions and the columns correspond to the variables, we create a definition for the dimensional matrix in Lean so we can ensure all dimensional matrices are of the same form.
\begin{code}
/*!\codeLink{DimensionalAnalysis/Basic.lean\#L269}!*/
/--Converts a list (tuple) of dimensions (the variables) into a matrix 
of exponent values-/
def dimensional_matrix {n : ℕ}  [Fintype B] (d : Fin n → dimension B E) (perm : Fin (Fintype.card B) → B) : Matrix (Fin (Fintype.card B)) (Fin n) E := Matrix.of.toFun (fun (a : Fin (Fintype.card B)) (i : Fin n) => d i (perm a))
\end{code}
This definition requires two fields: the list of variables, which has the type \textit{Fin n → dimension \(\alpha\) \(\gamma\)} meaning a tuple of n dimensions, and the permutation of the base dimensions. The permutation is not unique and is a way of picking a numerical order in which the base dimensions are indexed. By taking in a permutation, we can write a traditional matrix, like the one shown above. A possible permutation for \textit{SpatialTemporalSystem}, defined in a previous section, is:
\begin{code}
/-- Example permutation definition using the SpatialTemporal system -/
def SpatialTemporalSystemPerm 
| (0 : Fin 2) => SpatialTemporalSystem.Length2
| (1 : Fin 2) => SpatialTemporalSystem.Time2
\end{code}
The two parts of the Buckingham \(\pi\) Theorem determine how many dimensionless numbers, called Pi groups, can be formed and what those dimensionless numbers are. The number of pi groups that are possible is given by Equation \ref{Max Pi Groups}, where \(n\) is the number of parameters and \(k\) is the rank of the matrix. The rank of the matrix represents how many unique base dimensions describe the variables. 
\begin{equation}
    \label{Max Pi Groups}
    p = n - k
\end{equation}
In Lean, this is defined as:
\begin{code}
/*!\codeLink{DimensionalAnalysis/Basic.lean\#L274}!*/
/--Defnition of the number of dimensionless parameters possible from a 
list of dimensions-/
noncomputable def number_of_dimensionless_parameters {n : ℕ}  [Fintype B] (d : Fin n → dimension B E) (perm : Fin (Fintype.card B) → B) := n - Matrix.rank (dimensional_matrix d perm)
\end{code}
The rank of the dimensional matrix will normally be equal to the cardinality of the system in use. However, there are cases where this won't be true, specifically if a system contains a base dimension that isn't used. For example, if we consider finding the Pi groups for the set of variables: length and area, using \textit{system1}, the matrix will look like:
\begin{equation}
    \label{Dim Matrix Example 2}
    \begin{bmatrix}
    1 & 2 \\
    0 & 0 
    \end{bmatrix}
\end{equation}
Finding the form of the dimensionless parameters is done by finding the kernel of the dimensional matrix.
\begin{code}
/*!\codeLink{DimensionalAnalysis/Basic.lean\#L279}!*/
/--Defnition of the dimensionless parameters from a list of dimensions (not unique)-/
def dimensionless_numbers_matrix {n : ℕ}  [Fintype B] (d : Fin n → dimension B E) (perm : Fin (Fintype.card B) → B) := LinearMap.ker (Matrix.toLin' (dimensional_matrix d perm))
\end{code}

\section{Derivatives of Physical Variables}
\label{Derivative Section}
Derivatives are ubiquitous in engineering calculations and, here, we show the implementation of derivatives on a single variable in Lean. Like most operators on physical variables, the derivative operates on the value and the dimension separately. For a single variable function, the definition of the derivative is:
\begin{equation}
    \frac{df(x)}{dx} = \lim_{h \to 0} \frac{f(x+h)-f(x)}{h}
\end{equation}
Looking at the dimensions of this equation, we recognize that h must have the same dimension as x since h is added to x. Then, $f(x+h)$ must have the same dimension as $f(x)$. Therefore, we can simplify this equation and find that the dimension of the derivative is (recognizing that the limit does not change the dimension of the formula):
\begin{equation}
    \frac{df(x)}{dx} \,[=]\, \frac{f(x)}{x}
\end{equation}
When it comes to dimensions, the derivative acts just like division. 
\begin{code}
/*!\codeLink{DimensionalAnalysis/Basic.lean\#L133}!*/
/-- Definition of the derivative and integral operator for a single 
variable dimension function -/
def derivative (f : dimension B E → dimension B E) (b : dimension B E) : dimension B E := (f b)/b
def integral (f : dimension B E → dimension B E) (b : dimension B E) : dimension B E := (f b)*b
\end{code}
Then, for a physical variable function, the derivative is defined as:
\begin{code}
/*!\codeLink{PhysicalVariables/Basic.lean\#L66}!*/
/--Definition of the derivative for a single physical variable function -/
protected noncomputable def deriv {B : Type u} {V : Type v} [NontriviallyNormedField V] {d1 d2 : dimension B V} (f : PhysicalVariable d1 → PhysicalVariable d2)
(x : PhysicalVariable d1) : PhysicalVariable (d2/d1) :=
  let val' := deriv (PhysicalVariable.to_val_fun f) x.value
  ⟨val'⟩
\end{code}
This uses a function to convert a phyisical varaible function into a function of just values:
\begin{code}
/*!\codeLink{PhysicalVariables/Basic.lean\#L62}!*/
/-- Converts a physical variable function into a function of the value -/
protected def to_val_fun {B : Type u} {V : Type v} [Field V] {d1 d2 : dimension B V} (f : PhysicalVariable d1 → PhysicalVariable d2) : V → V
| a => (f ⟨a⟩).value
\end{code}

\end{document}